\documentclass[conference]{IEEEtran}
\IEEEoverridecommandlockouts
\usepackage{cite}
\usepackage{amsmath,amssymb,amsfonts}
\usepackage{algorithmic}
\usepackage{graphicx}
\usepackage{textcomp}
\usepackage{xcolor}
\def\BibTeX{{\rm B\kern-.05em{\sc i\kern-.025em b}\kern-.08em
    T\kern-.1667em\lower.7ex\hbox{E}\kern-.125emX}}

\usepackage[colorinlistoftodos,prependcaption,textsize=tiny]{todonotes}
\usepackage{comment}
\usepackage{csquotes}
\usepackage{hyperref}
\usepackage{tikz}
\usetikzlibrary{positioning, shadows, shapes.geometric, shapes.symbols, patterns}
\newcommand*\pointer[1]{\tikz[anchor=2mm]{\node[shape=circle,fill=black, text=white,scale=0.5] (char) {\textbf{#1}};}}

\usepackage{hyperref} 
\usepackage{todonotes}

\usepackage{graphicx}
\usepackage{subcaption}

\usepackage[scaled]{beramono}

\begin{document}

\title{Hawk: DevOps-driven Transparency and Accountability in Cloud Native Systems}

\author{\IEEEauthorblockA{Anonymous author(s)}}

\author{\IEEEauthorblockA{Elias Grünewald, Jannis Kiesel, Siar-Remzi Akbayin, and Frank Pallas\\
    \textit{Information Systems Engineering}, Technische Universität Berlin\\
    \{gruenewald, kiesel, s.akbayin, frank.pallas\}@tu-berlin.de}}

\maketitle

\begin{abstract}
    Transparency is one of the most important principles of modern privacy regulations, such as the GDPR or CCPA. To be compliant with such regulatory frameworks, data controllers must provide data subjects with precise information about the collection, processing, storage, and transfer of personal data. To do so, respective facts and details must be compiled and always kept up to date. In traditional, rather static system environments, this inventory (including details such as the purposes of processing or the storage duration for each system component) could be done manually. In current circumstances of agile, DevOps-driven, and cloud-native information systems engineering, however, such manual practices do not suit anymore, making it increasingly hard for data controllers to achieve regulatory compliance.

To allow for proper collection and maintenance of always up-to-date transparency information smoothly integrating into DevOps practices, we herein propose a set of novel approaches explicitly tailored to specific phases of the DevOps lifecycle most relevant in matters of privacy-related transparency and accountability at runtime: Release, Operation, and Monitoring. 
For each of these phases, we examine the specific challenges arising in determining the details of personal data processing, develop a distinct approach and provide respective proof of concept implementations that can easily be applied in cloud native systems. We also demonstrate how these components can be integrated with each other to establish transparency information comprising design- and runtime-elements. Furthermore, our experimental evaluation indicates reasonable overheads.
On this basis, data controllers can fulfill their regulatory transparency obligations in line with actual engineering practices. %

\end{abstract}


\section{Introduction}

Cloud native architectures enable data controllers to implement and operate scalable applications in modern, dynamic environments. These applications are modular and allow the implementation of decoupled services, which has manifold advantages, but makes the application of a fully comprehensive data protection concept difficult. Privacy regulations -- such as the European General Data Protection Regulation (GDPR) \cite{gdpr} or the California Consumer Privacy Act (CCPA) \cite{ccpa} -- raise obligations and constraints that must be complied with. %

In matters of transparency, which we mainly address herein, this means that data controllers must inform data subjects (such as their customers) with information about the details of personal data processing to allow for well-informed decisions. To achieve accountability, in turn, data controllers need to be able to demonstrate their technical and organizational compliance measures, which is regularly met through keeping records of processing activities (RoPAs) \cite{edpbPbD}. For cloud native and DevOps-focused systems engineering, this proves increasingly problematic, as the underlying facts and technical givens change at high frequency and are shaped by independent teams \cite{intercept}. Technical tools allowing to collect and maintain relevant information in cloud native systems architectures and specifically tailored to the regulatory requirements regarding personal data thus become indispensable for achieving compliance. %

Meanwhile, to be applicable and adopted in practice, technical tools must integrate well into broadly established paradigms and practices of modern systems engineering. Even though transparency enhancing technologies (TETs) \cite{zimmermann2015}%
have been proposed in a wide variety of forms, these do so far not explicitly address the specific challenges of cloud native environments. Although some work has been done in projects like A4Cloud \cite{pearson2012accountability}, we identify several open challenges. In particular, developers are ill-equipped with tools that integrate with containerized microservice architectures, constantly exchanging (personal) data through APIs \cite{gruenewald2021tira}. Moreover, we build on the premise that software developers generally do not have a well-developed knowledge of the legal regulations.%
Consequently, we need to address this challenge with easy-to-adopt approaches and their technical implementation in modern systems following the DevPrivOps lifecycle \cite{gruenewald2022devprivops}.

We address these multifaceted challenges with the following contributions. All of these come with a general approach and a prototypical implementation enabling data controllers to

\begin{itemize}
    
    \item perform continuous deployments, e.g. canary releases, that can detect and highlight differences in the personal data flows between versions of a microservice, integrated into existing deployment setups (\textit{Hawk Release}), and
    \item label and track personal data flows between running microservices with a service mesh extension and a developer-friendly library (\textit{Hawk Operate}), and
    \item collect, aggregate, expose, and visualize metrics about these personal data flows %
    (\textit{Hawk Monitor}). 
\end{itemize}

\noindent Additionally, we provide
\begin{itemize}
    \item a preliminary performance evaluation illustrating suitable overheads in a realistic setting.
\end{itemize}

This paper includes the following sections: related work (sec.~\ref{sec:related-work}), general approach (sec.~\ref{sec:general-approach}), and phase-specific contributions: \textit{Hawk Release} (sec.~\ref{sec:release}), \textit{Operate} (sec.~\ref{sec:operate}), and \textit{Monitor} (sec.~\ref{sec:monitor}). We evaluate our findings (sec.~\ref{sec:evaluation}) and provide a discussion (sec.~\ref{sec:discussion}). %

\section{Background \& Related Work}\label{sec:related-work}

Our work emanates from the interplay of three different areas of research and practice, to be laid out in brief. %

\paragraph{Cloud Native, Microservices, and Service Mesh} \label{ssec:cloud-native}
To begin with, we define cloud native systems following the Cloud Native Computing Foundation (CNCF) \cite{cncf}. They refer to scalable applications, typically running in public cloud settings, %
built with containerized microservices. Microservices are comparatively small units with a dedicated business function (bounded context) and they are (optimally) independently developed and operated by small developer teams \cite{dragoni2017microservices}. In cloud native systems, these services are then dynamically orchestrated to optimize communication and resource utilization \cite{cloudnative}. Since these systems quickly grow in complexity, several best practices evolved to better manage, observe, and secure their behavior. One popular approach is the integration of a service mesh \cite{calcote2019istio}. In such settings, each microservice is equipped with an infrastructure layer that does not impose the internal implementation \cite{li2019service}. Most approaches, such as Istio, add dedicated data and control planes within the container orchestration engine. Within the data plane, all traffic from, to, and between microservices is conditionally manipulated or routed to enable load balancing, authentication and authorization, or observability. %
Service meshes have primarily been employed to guarantee security-related policies \cite{chandramouli2020building}. %

\paragraph{Privacy, Transparency, and Personal Data}

Privacy regulations raise obligations for data controllers to inform data subjects and supervisory authorities. %
Art.~12~ff. of the GDPR, for instance, require data controllers to inform data subjects about which personal data are collected for which purpose(s), storage periods, possibly existing transfers and recipients, etc. \cite{art29transparency, tilt} %
This information is understood as an indispensable precondition for data subjects to act in a well-informed manner and to execute their rights \cite{art29transparency}. %

Transparency information is typically provided through written privacy policies. Such policies have been criticized for being overly long and written in highly legalese terms \cite{ReidenbergDisagreeablePrivacyPolicies2015, Linden2018}, leading to data subjects barely reading, let alone understanding them \cite{rudolph2018}. Proposals for addressing respective shortcomings range from \enquote{layered} textual representations \cite{art29transparency} over more versatile modalities of information provision and presentation \cite{gruenewaldEnablingVersatile} such as chatbots \cite{pribots} or visual icons \cite{holtz, habib2021icons} to formal, machine-readable representations \cite{gerl2019, tilt} facilitating a broad variety of possible interfaces.
Under today's givens of cloud-native DevOps practices (cf.~\cite{bass2015devops}), %
manual bookkeeping about which system components process, or store what personal data where, for which purpose, and how long is neither reasonable nor feasible anymore. Instead, automated technical approaches and tools are needed. %

\paragraph{DevOps-driven Transparency Technologies} \label{ssec:devprivops} 

In order to be properly provided, transparency information (and the technical givens it is meant to represent) must first be compiled and continuously maintained throughout the whole lifecycle of the system(s) collecting and processing personal data (see, e.g., \cite{gurses_privacy_agile, kostova2020privacy, sion2020neverending, gruenewald2022devprivops}). Traditionally, this has been done through comparably manual inventory procedures, carefully and consciously collecting \enquote{records of processing activities} (cf. Art. 30 GDPR) \cite{cnil}. %
Technical approaches proposed in this regard so far include tools for
automated (privacy) threat assessment in continuous integration pipelines \cite{sion2021automated},
tracking selected data flows in cloud environments \cite{data-track2016, kunz2020tracking},
decentralized transparency and consent logging \cite{havur2020greater}, and 
several proposals using information flow control \cite{pasquier2016, apiclarity}. However, they do not cover all the legally required transparency information (and are thus not capable of achieving compliance) and lack explicit alignment to DevOps practices and integration into cloud native approaches.

\section{General Approach}\label{sec:general-approach}

We provide novel contributions
explicitly reflecting the DevOps phases \cite{bass2015devops}.
In the following, we scope the phases particularly relevant for \textit{runtime compliance} challenges: \textit{Release}, \textit{Operate}, and \textit{Monitor}.
Henceforth, we identify the typical legal and engineering challenges for each of the three phases (DevPrivOps) \cite{gruenewald2022devprivops}. On this basis, we propose a general approach and a prototypical implementation to tackle them. Since all of them are interlinked through the continuous development practice, we also investigate the interactions of our components. %
We publish all of our prototypes as open-source software.%
\footnote{\url{https://github.com/PrivacyEngineering/hawk}} and aim for the following design goals:

\textit{Low integration effort:} Our approach should be easily integrable to microservice infrastructures without burdensome implementation efforts, in line with the requirements listed for similar endeavors of privacy engineering \cite{gruenewald2021tira, redCastle}

\textit{Manageable performance impact:} Since regulatory compliance is one of sometimes conflicting design goals with other system properties \cite{sun2020security}, such as scalability, we need to evaluate the actual performance overheads with experiments. This also allows for an evidence-based decision regarding the appropriateness according to Art.~25(2)~GDPR.

\textit{Regulatory expressiveness:} To be practical, adherence to the transparency and accountability principles from the GDPR with the required expressiveness (i.e., capturing all relevant information) \cite{tilt} is crucial from a regulatory standpoint.

\label{ssec:running-example}

We use the SockShop \cite{sockshop} as a running microservice example. It has repeatedly been used for manifold systems engineering demonstrations and features some realistic business functionalities relating to the processing of personal data through an online shop scenario.
We deployed the system %
in a cluster on the Google Kubernetes Engine.\footnote{\url{https://github.com/PrivacyEngineering/hawk-sockshop}} %

\section{Hawk release}\label{sec:release}

\begin{figure}[t]
    \centering
    \includegraphics[width=1.0\linewidth]{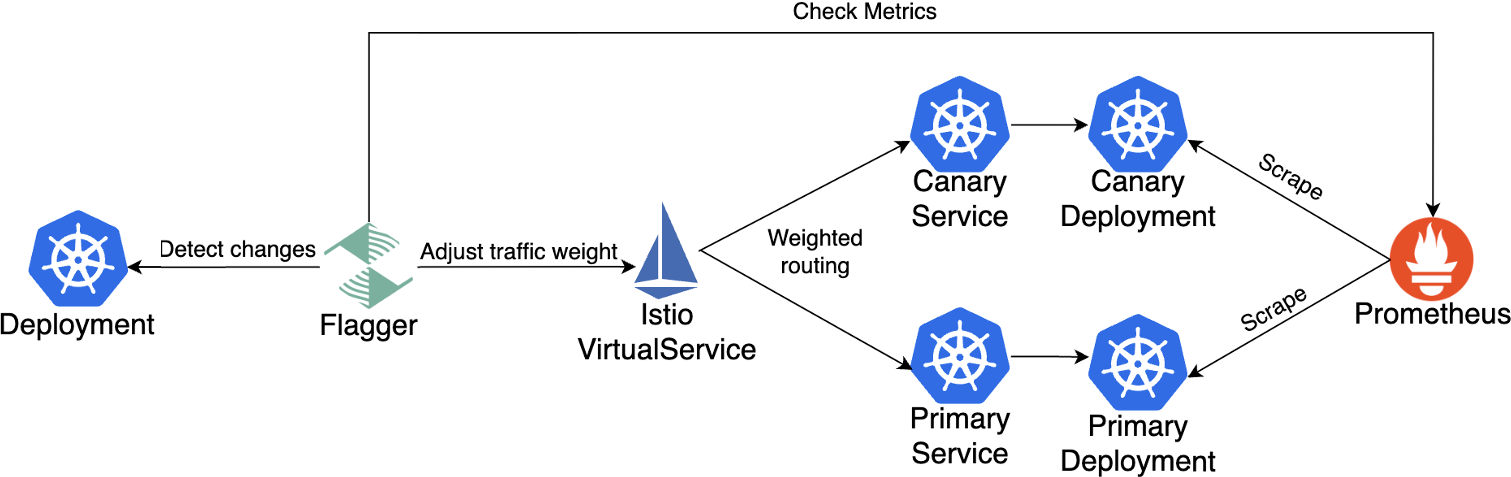}
    \caption{Canary release process.}
    \label{fig:release}
\end{figure}

In the following, we consider the particular challenges, our approach, and implementation regarding the \textit{Release} phase.%

\paragraph{Challenges}
Releasing a new version of a microservice to the production environment comes with several challenges. Through the DevOps practice, release cycles become comparably short, from months in a monolith to minutes in a well-modularized architecture. This is often beneficial from a business perspective to support better features \cite{devopsRingu2016}. However, this is oftentimes not compatible with in-depth privacy checks. Data controllers need to make sure that, e.g., the processed categories of personal data, their purpose specification, or enacted third-country transfers do not change without proper notice and appropriate technical-organizational measures to ensure full transparency and accountability. Furthermore, the GDPR obligates a risk-based approach. Consequently, this results in adequate risk analyses and estimations of severance in case of data breaches before any processing takes place. Without any DevOps-driven approach, this is performed through manual data protection impact assessments (DPIAs) which involve legal and developer expertise. However, these cannot factor in any runtime behavior resulting from the actual integration of an updated service (e.g., higher traffic volumes than expected). To meet the release pace and simultaneous risk considerations, we propose to introduce respective automated checks before the deployment.  

\paragraph{Approach}
Following the risk-based approach, we propose to deliver the new version of a service progressively -- thus, minimizing the risk for affected data subjects -- while checking certain configurable privacy metrics. These shall include common transparency information elements \cite{tilt}. The new version of a microservice, in which possible negative impacts might be detected, can then be rolled out completely or rolled back to stop the undesired effects. In particular, we propose a canary release strategy, separated into the four phases %
\cite{rapidErnst2019}, which we describe below. %

\paragraph{Implementation}
We briefly summarize our proof-of-concept implementation along the four phases mentioned before.\footnote{\url{https://github.com/PrivacyEngineering/hawk-release}}
Our setup follows pull-based GitOps practices. Hence, all necessary infrastructure is declaratively defined. We use the Flux toolkit, including an operator component that periodically checks on changes in the infrastructure to synchronize it with the running cluster. We depict the interplay of the core components in figure~\ref{fig:release}.

\textit{Deployment} Through the Flux \textit{Source}, \textit{Kustomize}, \textit{Helm}, \textit{Notification}, and \textit{Image Automation} controllers, we set up a full-fledged and highly automated continuous deployment pipeline, which can be applied to all running services. These controllers also manage deployments, secrets, and custom resource definitions necessary. The complete process is triggered by infrastructure-related changes for individual microservices in the Git-based version control system (VCS). Consequently, Flux applies changes that have been pushed to the VCS automatically to the cluster. Likewise, the changes on the cluster are detected by Flagger, which starts the canary release process automatically. Flagger supports different release strategies by providing a \texttt{Canary} custom resource definition in which the release process can be configured. 

\textit{Load Shifting} To realize the load shifting, Flagger provides the integration of different service mesh technologies. We opted for Istio which provides a custom resource called \texttt{VirtualService} to set load routing rules, including the traffic distribution between different service versions. This custom resource gets generated automatically %
based on the specific configuration of the \texttt{Canary} custom resource.

\textit{Observation} Flagger provides a custom resource called \texttt{MetricTemplate} that can be used to query metrics from different providers and refer to them in the \texttt{Canary} object. These metrics can then be used as a threshold for the promotion of a new version. We define specific custom Prometheus metrics regarding regulatory transparency (such as uncategorized personal data indicators, see below) %
to assure that the new version does not conflict with pre-defined requirements. %

\textit{Cleanup} After a successful canary release, Flagger scales the old version of the microservice automatically to zero (i.e., removes the instances) and marks the new version as \textit{primary}. %

\textit{Metrics} To capture transparency information, we define custom Istio metrics collected through the Envoy proxies within the service mesh. In particular, we use the three types of Prometheus metrics: \texttt{Counter}, that can be increased or reset, \texttt{Gauge} that can be increased and decreased, and \texttt{Histogram} which samples observations in pre-defined buckets. We capture key-value pairs following the legal transparency information elements, that can then be queried through the Prometheus Query Language (PromQL). %

We implemented three illustrative examples within our prototype. First, we created an Envoy filter and a corresponding \texttt{Counter} metric for capturing third-country transfers (cf.~Art.~44ff.~GDPR). Doing so, we resolve IP addresses for geolocations %
and further classify them to EU and Non-EU countries following the regulatory provisions.
Second, we captured the interplay between two specific microservices. In this example, we assume any \texttt{POST} request to one of them indicates a payment process. Following the purpose limitation principle (Art.~5(1b)~GDPR), personal data can only be processed for a specified, explicit and legitimate purpose \cite{Finck_Biega_2021}. With such a metric, we can observe any service sending data to the payment microservice without an explicit purpose statement, indicating a potential violation. %
The metric records which services made how many requests, also allowing to quantify certain behavior changed through the new version of a service. %
Third, we check for metrics from the \textit{Hawk Operate} service (cf. sec.~\ref{sec:operate}).   
All in all, we present a general approach for decreasing the likelihood of missing out transparency requirements and decrease manual efforts for records keeping. %

\section{Hawk Operate}\label{sec:operate}
In this section, we present the particular challenges, our approach, and implementation regarding the \textit{Operate} stage.%

\begin{figure*}
  \begin{minipage}[b]{0.34\textwidth}
    \centering
    \includegraphics[width=\linewidth]{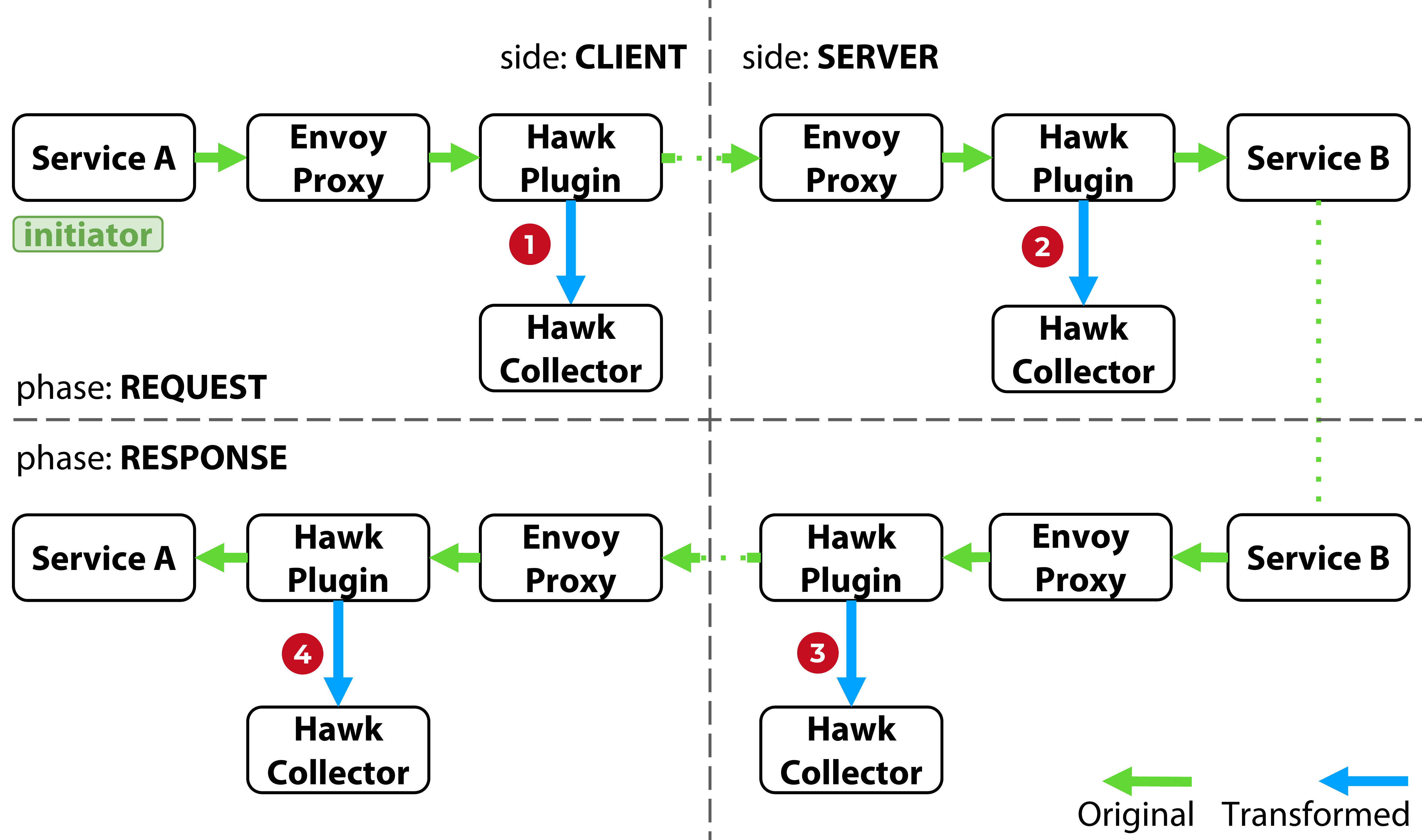}
    \caption{Message flow.}
    \label{fig:message-flow}
  \end{minipage}
  \hfill
  \begin{minipage}[b]{0.64\textwidth}
    \centering
	\includegraphics[trim=0.0cm 14.5cm 0.0cm 0.0cm, width=\linewidth]{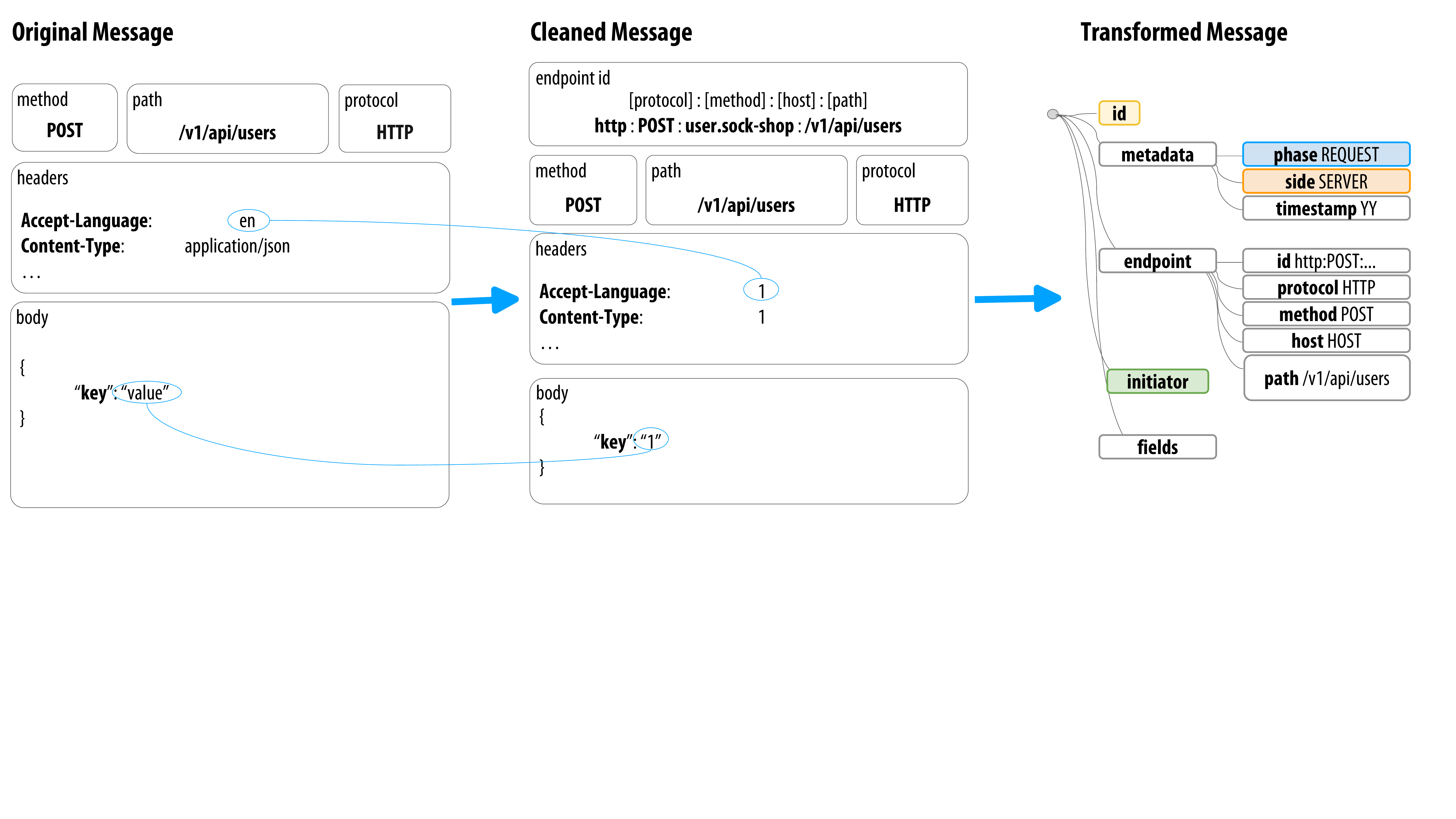}
	\caption{Message transformation.}
	\label{fig:messageTransformationHawkPlugin}
  \end{minipage}
\end{figure*}

\paragraph{Challenges}
Data controllers face difficulties evaluating all changes regarding personal data flows \textit{at runtime}, especially considering fast-paced release cycles (cf.~\ref{sec:release}). %
Particularly, the consistent labelling of data flows involving personal data (and, thus, separating from non-personal data) needs to occur automatically and should not significantly impede the system performance. %
Clearly, the process of identifying personal data as such, cannot be fully automated, since the legal classification is based on numerous criteria (see also \cite{finck2020they, gruenewald2022teiresias}).
Another challenge is to capture all incoming and outgoing requests of each microservice, potentially processing personal data. %
In addition, it is error-prone for developers without expertise in privacy regulation, to inventory the relevant records. Therefore, we argue to capture relevant transparency information not within an application, since this might neither be feasible for proprietary services.%

\paragraph{Approach}

Tracking requests containing personal data could be achieved through feature-rich libraries and labelling mechanisms for each microservice. %
However, this would imply the necessary implementation for numerous programming languages and frameworks, notwithstanding the additional integration overheads for each team. We solve this scalability problem through the integration of a service mesh, and a corresponding proxy extension. Meanwhile, we provide an additional library for interoperability in cases a proxy cannot be used.%

Being able to capture the relevant traffic, we propose a configurable labelling mechanism without storing or transmitting actual personal data items. For this reason, we present an approach for selective information extraction and transformation of exchanged messages through path identifiers, a labelling data structure, and a suitable configuration dashboard.%

\paragraph{Implementation}

Integrating a service mesh allows for directing traffic and observability in a Kubernetes cluster. This allows for capturing traffic between origin (i.e., initiator or client) and destination (or server) services exchanging messages. As mentioned in section~\ref{sec:release}, we choose the Istio service mesh with Envoy proxies. Therefore, our implementation can also be adapted for other service mesh technologies, such as AWS App Mesh, since many rely on the same proxy.

First, we extended the proxy through a custom layer~7 HTTP filter based on the Web Assembly (WASM) sandbox written in TinyGo (\textit{Hawk plugin)}.\footnote{\url{https://github.com/PrivacyEngineering/hawk-envoy-plugin}} This results in a common message flow and interpretation pattern as depicted in figure~\ref{fig:message-flow}. Service A initiates an HTTP request to Service B with a JSON message (potentially containing personal data). This message is routed through the Proxy, and gets intercepted by the \textit{Hawk plugin}. Meanwhile, the Proxy will generate a unique request ID and append it to the message header. The extension filter will transform~\pointer{1} the message to extract the required information to be sent to the \textit{Hawk Collector}. Likewise, the message arrives at the destination Pod, and it is intercepted by the proxy. The message is again transformed~\pointer{2} to be collected. As the destination service is reached, the response can be generated, and the message flow resumes in the same way back to the initiator, including two more~\pointer{3},~\pointer{4} transformations and collections. In total, the \textit{Hawk Plugin} processes four messages, one for each stage. Each time, different kinds of meta-data are collected, as depicted in figure~\ref{fig:messageTransformationHawkPlugin}. Our \textit{Hawk Collector} component expects messages, which contain the request ID, metadata depending on the phase (request or response), the side (client or server), the current timestamp, the HTTP meta-data (protocol, method, host, and path), and some auxiliary fields (such as the precise endpoint). Moreover, the collected data fields from the header and payload are expected after a cleaning and aggregating transformation.
In the cleaning step, to minimize the amount of personal data collected, we only keep the existing properties, but not their values.
Thereby, only the structure -- represented as JSON path expression --, not the actual content, is kept. For instance, a JSON payload that contains the property $k$, results in a JSON path expression \texttt{\$.k}.

Second, we implemented the \textit{Hawk Collector} written in Go. It receives the transformed messages through a RabbitMQ message queue (for scalability and delivery guarantees). It catches any invalid messages for further inspection (audit trail). The collected information is then transferred to the \textit{Hawk Core} service (see below).

Third, we provide a library to demonstrate that -- apart from using the extended Envoy plugin, which e.g., would not work for encrypted communication or other services which can not be intercepted -- sending transparency information to the \textit{Hawk service} directly is also supported within our framework.\footnote{\url{https://github.com/PrivacyEngineering/hawk-integration-java/}} In particular, we implemented a Java/Kotlin integration for the Spring framework. This works through code annotations (i.e., \texttt{@EnableDataUsageTracing}), which then automatically clean and transform the relevant messages through the filter interface of Spring Web. We also provide feature-rich integration settings for different kinds of transparency information to be collected along the regulatory givens. 

Finally, we implemented the \textit{Hawk Core} service, which receives the collected messages for aggregation, storage, and to enable the creation of subsequent analysis and reports.\footnote{\url{https://github.com/PrivacyEngineering/hawk-service/}} %
Since the \textit{Collector} provides the processed fields, these need to be labelled as actually referring to personal data (personal data indicators).
\textit{Hawk Core} enables the definition of personal data indicators following the legal provisions and offers a semi-automatic mapping approach of these labels to the emitted messages from individual endpoints. For instance, these can also be enriched with transparency information that cannot be extracted out of messages between services. These include, for instance, field names, purpose, and description, whether they are personal data under special category (Art.~9~GDPR), %
\textit{et cetera} \cite{tilt}. Based on all the received data, it prepares the necessary input for RoPAs, and metrics for privacy-aware releases (cf. section~\ref{sec:release}).  

\textit{Hawk Operate} enables the collection of individual and batched formatted messages, as well as CRUD mappings of formatted messages to specific fields. The integration into large-scale systems, requires the \textit{Collector} and \textit{Core} to scale accordingly. Horizontal scaling is supported by stateless design, and database interaction is implemented through an object relational mapper, that allows the utilization of any relational column store to meet the insert-most workload with some time-based analytical queries. %

\section{Hawk monitor}\label{sec:monitor}
\noindent We continue with the \textit{Monitor} stage.

\paragraph{Challenges}

The necessary information for effective transparency and accountability needs to be monitored continuously. Related work on static transparency information, for instance in the form of privacy policies, can neither comprehensively present the complex data flows in a cloud native system nor can it be updated without immense manual efforts. Meanwhile, the GDPR also obligates data controllers to perform runtime monitoring \cite{sion-landuyt2021runtime}. However, related work does only partially address this challenge, e.g., for data minimization \cite{pinisetty2018monitoring} or security reasons \cite{casola2015security, sion2021automated}. %
Therefore, transparency-focused monitoring is likely to be legally required (Art.~12--14,~25,~30~GDPR) \cite{runtime2019}. In complex systems, one of the core challenges is to cover as many data flows as possible. In addition, the collected information needs to be aggregated to also gain insights over time, since the functionality and the resulting processing activities change constantly. In addition, microservices may communicate with external services that (sometimes unexpectably) change APIs. %

\paragraph{Approach}
We propose the general \textit{Hawk Monitor} approach, which builds upon the information collected through \textit{Hawk Operate}, and eventually displays aggregated privacy data flows between services through configurable dashboards.%
Data controllers can monitor privacy-related data exchanges between services and get insights through the service mesh extension and/or the programming library. Some characteristics, such as the data volume and most-used endpoints, can be inferred directly from the monitored traffic. These might be important data points for risk-based data protection impact assessments. %
We also propose a configuration dashboard for managing the legal bases for processing, their purpose specifications, and other attributes, according to the applicable privacy regulations.%

\paragraph{Implementation}
The \textit{Hawk Core} service retrieves runtime information using mapping templates for endpoint annotation. These are common labels for collected endpoints and their respective personal data indicators. Once all relevant fields are annotated, users can start investigating how exactly, according to the regulations, (personal) data are being processed at the moment of inspection or over an arbitrary period of time. \textit{Hawk Monitor}\footnote{\url{https://github.com/PrivacyEngineering/hawk-monitor}} offers versatile configurations, visualization, aggregation, and querying capabilities. %
Features include a configuration dashboard, a dynamic service graph, and tabular summaries.
First, the configuration dashboard summarizes the currently available mappings, set personal data indicators, and templates for future annotations. We assume this annotation to be joint work by the developers and legal experts. %
Second, the dynamic service graph shows the different running services and their current number of API requests. The panel highlights the data flows through animated packages and serves as a general overview of all processing activities. Third, we provide tabular summaries, for instance, the requests per microservice, an overview of initiating message flows (cf.~\ref{sec:operate}), requests per endpoint (e.g., filtered per purpose specification), and others. These can be filtered for specific data fields. Hence, the data controller can precisely check where (even in large-scale infrastructures) a particular data item is being processed. Moreover, the underlying data allows for different types of range queries or aggregations.%

Altogether, \textit{Hawk Monitor} delivers valuable indicators in matters of compliance. Notably, the information presented can be used to observe the system and introduced changes at runtime. With this novel approach, we go far beyond related work that only presents static information \cite{angulo2015usable}.

\section{Preliminary evaluation}\label{sec:evaluation}

In this section, we summarize some performance benchmarking results. We deliberately emphasize their preliminary character, since we did not perform any optimizations so far. %
We conducted our experiments with the \textit{SockShop} introduced in section~\ref{ssec:running-example}. We provide our raw data for repeatability.\footnote{\url{https://github.com/PrivacyEngineering/hawk-benchmark}}

\begin{figure*}
  \begin{minipage}[b]{0.33\textwidth}
  \centering
    \includegraphics[height=0.7\linewidth]{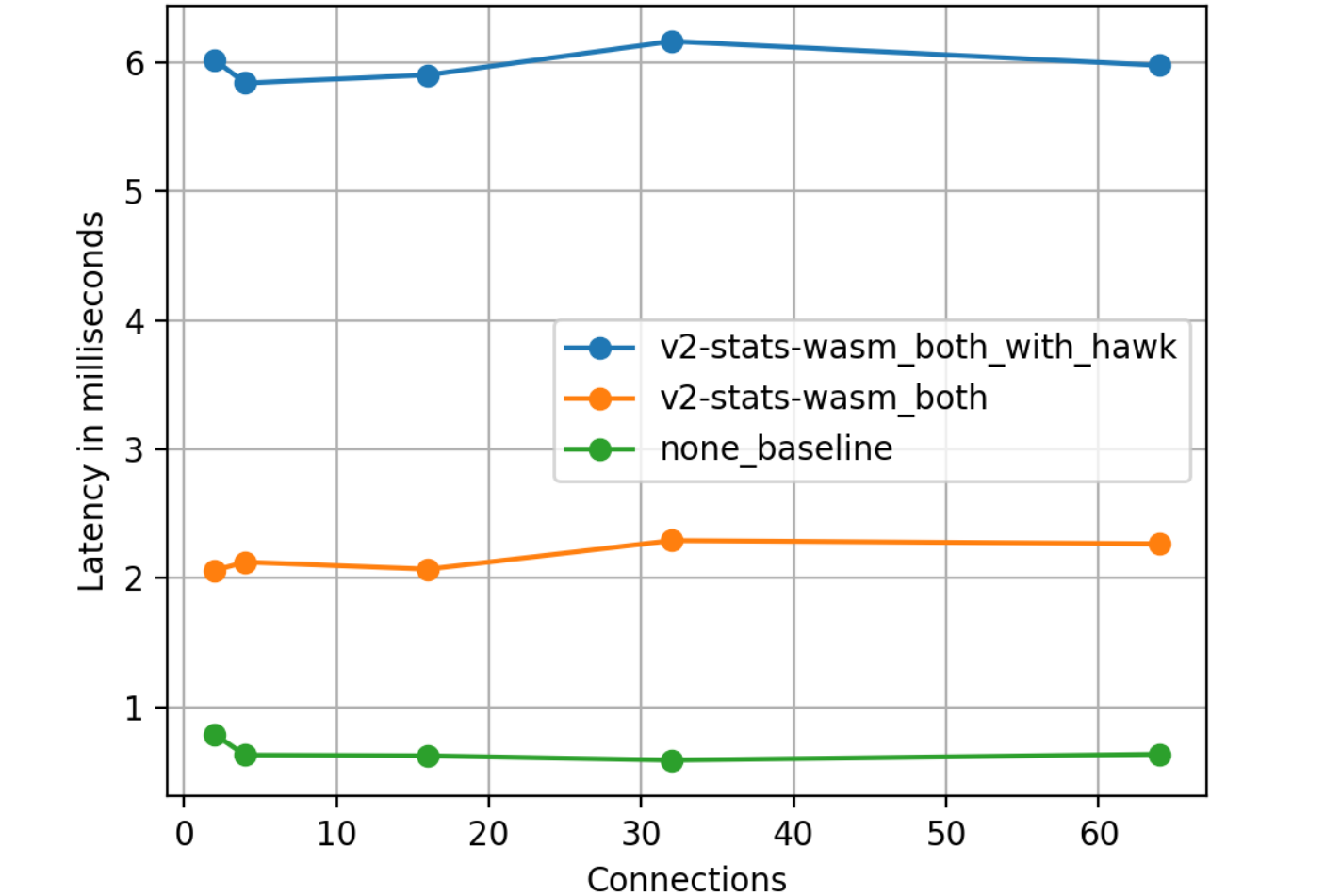}
    \caption{Latency comparison.} %
    \label{fig:filter_latency}
  \end{minipage}
  \begin{minipage}[b]{0.33\textwidth}
    \centering
    \includegraphics[height=0.7\linewidth]{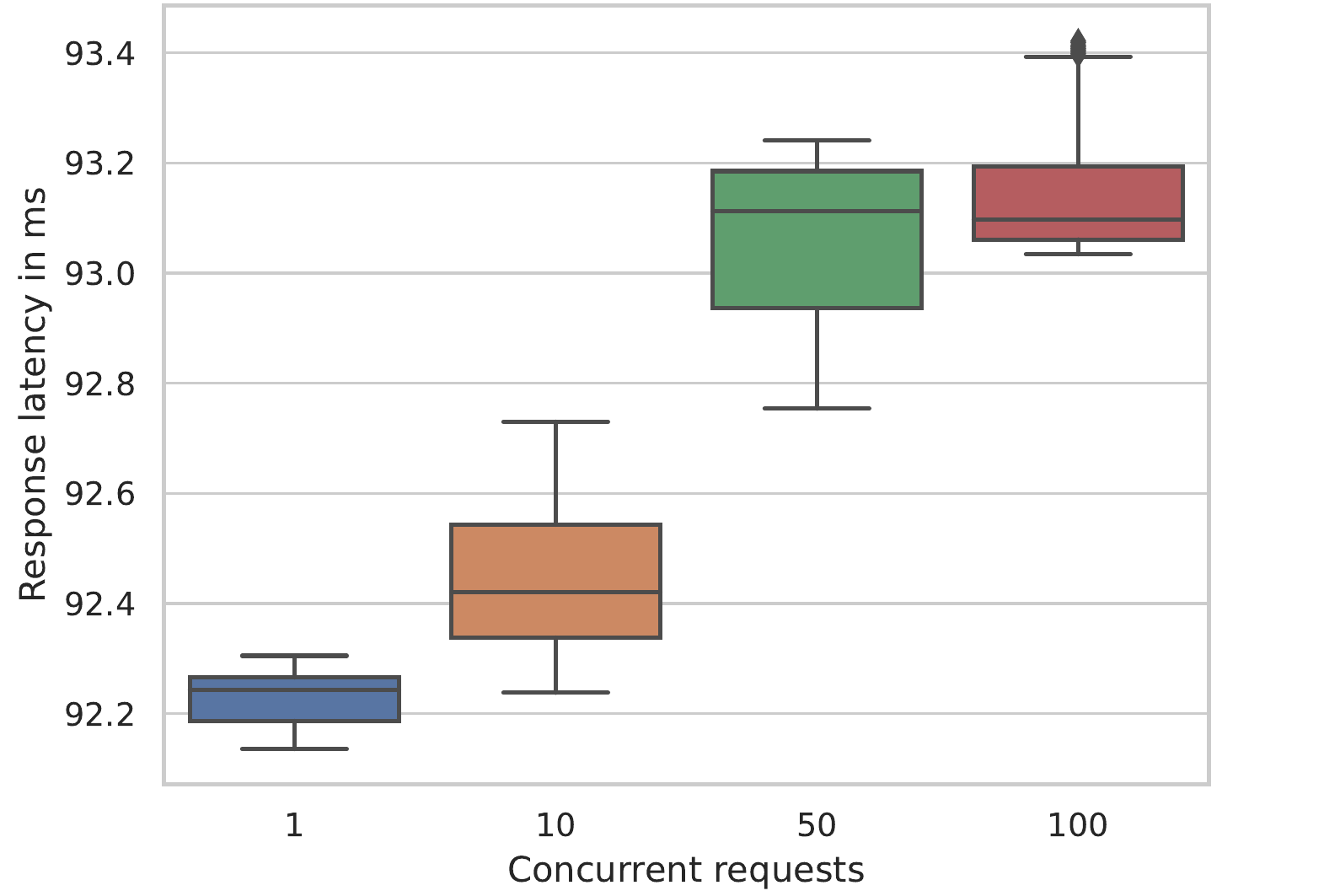}
    \caption{Collector latency.} %
    \label{fig:collector_latency}
  \end{minipage}
  \begin{minipage}[b]{0.33\textwidth}
    \centering
    \includegraphics[height=0.7\linewidth]{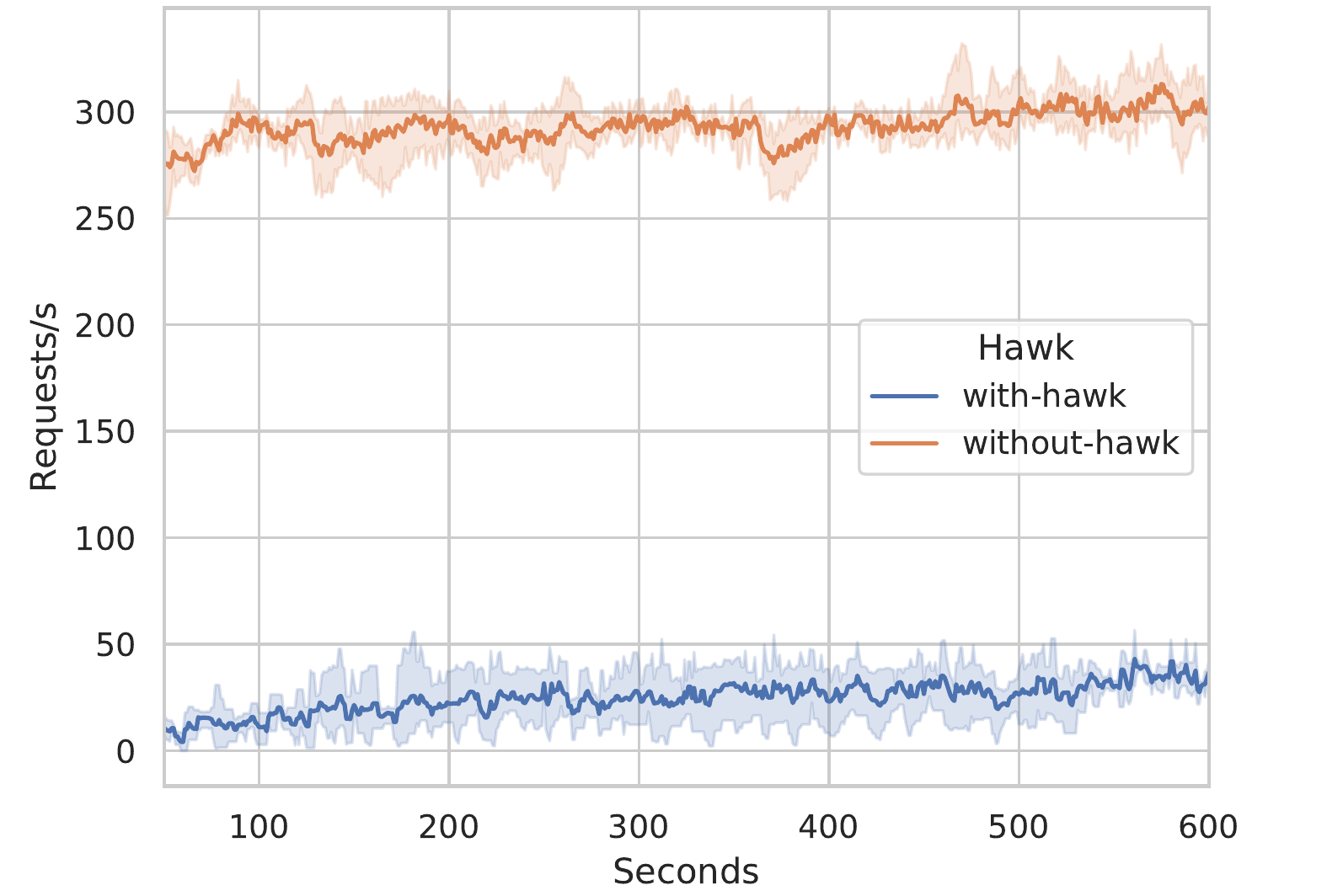}
    \caption{Throughput.} %
    \label{fig:throughput_100_users}
  \end{minipage}
\end{figure*}

For all benchmarks, we created a cluster running on the Google Kubernetes Engine with three \textit{n1-standard-8} nodes, 24~vCPUs, and 90~GiB total memory in the \textit{us-central1-c} availability zone. Furthermore, we deployed Istio version 1.18-alpha and the \textit{default} %
profile as base configurations. 
The benchmarking clients were placed in the same virtual private network to reduce network-related delays.

First, we measure the overhead in the message flow introduced by the \textit{Hawk plugin}. %
We utilize the commonly used benchmarking tool of Istio \cite{istio_bench}
and deploy two Fortio \cite{fortio} pods (one client and server) which communicate over HTTP using mTLS. The client pod performs HTTP requests with a 1~KB payload. CPU and network utilization are monitored through the cloud platform for a fair benchmark. 

We started the provided performance tests with the v2 telemetry stats WASM filter activated, since the \textit{Hawk plugin} filters are also of the same kind. The tests used Nighthawk \cite{nighthawk} for the load generation and returned the latency for up to 64 client connections generating 100 req/s. All tests ran 240~s per connection number variation. The results for the 90\textsuperscript{th} percentile are illustrated in figure~\ref{fig:filter_latency}. As a result, the latency triples in comparison to the default v2 telemetry WASM filter if the \textit{Hawk plugin} is applied. However, this highly depends on the response body size, which varies in real systems. We acknowledge reasonable overheads for many scenarios, and aim to decrease the overheads through performance optimization, e.g., through caching. 

Second, we ran several experiments benchmarking against the \textit{Hawk Collector}. We used the Locust%
load generator to measure both latency and throughput, as depicted in the figures~\ref{fig:collector_latency}~and~\ref{fig:throughput_100_users}. The latency measured only increases minimally as the number of simultaneous requests grows from 1 to 100. This satisfies the criteria of reasonable overheads. Measuring the throughput with 100 concurrent users (repeated 3 times) yields significant performance decreases up to one order of magnitude. Given the manifold advantages, this still might be acceptable for many risky scenarios, %
while further optimization, such as sampling and event-triggered activation of all monitoring features shall be explored.

\section{Discussion, Future Work, and Conclusion}\label{sec:discussion}
\noindent In this last section, we briefly discuss our findings.%

Regarding the initial design goal of low integration effort, we presented the \textit{Hawk Operate} approach based on a service mesh, which can easily be integrated without changing the application code. The performance impacts have been estimated through our preliminary evaluation. Further optimizations seem to be promising, while the approach already unfolds feasible for the scenario at hand. %

Up until now, \textit{Hawk Release} supports canary releases. In future work, additional deployment strategies should be explored for their applicability. These could include A/B tests, Blue/Green deployments, and others. As a whole, these could also be used to implement additional safeguards apart from threshold checks against the metrics already presented. In fact, such pipelines could be designed with more feature-rich sets of rules, multi-cloud support, and verbose reporting.

Moreover, the \textit{Hawk} framework could profit from additional input sources and output formats. %
We think of, inter alia, data loss prevention systems \cite{gruenewald2022teiresias}. Distributed tracing tools could provide additional valuable insights. Likewise, additional proxy extensions or libraries could be covered.

In summary, our DevOps-based techniques significantly enhance the capacity of data controllers to gain a comprehensive overview of their systems (to adopt a \textit{bird's-eye view}), while simultaneously enabling them to access detailed transparency data regarding individual requests. We also emphasize the new collaboration and communication workflows for technical and legal experts within a data controller. These diverse options serve as a foundation for various research inquiries from a wider privacy engineering perspective.

\section*{Acknowledgments}
\begin{small}
We thank Paul~Bennert, Juan~M.~Goyes~Coral, Zlatina~Metodieva, Paskal~Paesler, and Piotr~Witkowski, who contributed to the initial implementation within the scope of a prototyping course at TU~Berlin. We also thank Louis~Loechel for supporting the preliminary evaluation.

The work behind this paper was funded within the project TOUCAN\footnote{\url{https://tu.berlin/ise/toucan}}, supported under grant no. 01IS17052 by funds of the German Federal Ministry of Education and Research (BMBF) under the Software Campus 2.0 (TU Berlin) program.
\end{small}

\bibliographystyle{myieeetran} %
\bibliography{references}

\begin{thebibliography}{10}
\providecommand{\url}[1]{#1}
\csname url@samestyle\endcsname
\providecommand{\newblock}{\relax}
\providecommand{\bibinfo}[2]{#2}
\providecommand{\BIBentrySTDinterwordspacing}{\spaceskip=0pt\relax}
\providecommand{\BIBentryALTinterwordstretchfactor}{4}
\providecommand{\BIBentryALTinterwordspacing}{\spaceskip=\fontdimen2\font plus
\BIBentryALTinterwordstretchfactor\fontdimen3\font minus
  \fontdimen4\font\relax}
\providecommand{\BIBforeignlanguage}[2]{{%
\expandafter\ifx\csname l@#1\endcsname\relax
\typeout{** WARNING: IEEEtran.bst: No hyphenation pattern has been}%
\typeout{** loaded for the language `#1'. Using the pattern for}%
\typeout{** the default language instead.}%
\else
\language=\csname l@#1\endcsname
\fi
#2}}
\providecommand{\BIBdecl}{\relax}
\BIBdecl

\bibitem{gdpr}
{European Parliament and Council of the European Union}, ``Regulation ({EU})
  2016/679 of 27 {April} 2016 on the protection of natural persons with regard
  to the processing of personal data and on the free movement of such data, and
  repealing {Directive} 95/46/ec ({General Data Protection Regulation}),''
  2018.

\bibitem{ccpa}
{California Civil Code}, ``California consumer privacy act ({CCPA}),'' 2018.

\bibitem{edpbPbD}
\BIBentryALTinterwordspacing
{European Data Protection Board}, ``Guidelines 4/2019 on article 25 data
  protection by design and by default,'' 2019. [Online]. Available:
  \url{https://edpb.europa.eu/our-work-tools/our-documents/guidelines/guidelines-42019-article-25-data-protection-design-and_en}
\BIBentrySTDinterwordspacing

\bibitem{intercept}
\BIBentryALTinterwordspacing
S.~Biddle. (2022) Facebook engineers: We have no idea where we keep all your
  personal data. [Online]. Available:
  \url{https://theintercept.com/2022/09/07/facebook-personal-data-no-accountability/}
\BIBentrySTDinterwordspacing

\bibitem{zimmermann2015}
\BIBentryALTinterwordspacing
C.~Zimmermann, ``A categorization of transparency-enhancing technologies,''
  arXiv, 2015. [Online]. Available: \url{https://arxiv.org/abs/1507.04914}
\BIBentrySTDinterwordspacing

\bibitem{pearson2012accountability}
S.~Pearson, V.~Tountopoulos, D.~Catteddu, M.~Südholt, R.~Molva, C.~Reich,
  S.~Fischer-Hübner, C.~Millard, V.~Lotz, M.~G. Jaatun \emph{et~al.},
  ``Accountability for cloud and other future internet services,'' in \emph{4th
  IEEE international conference on cloud computing technology and science
  proceedings}.\hskip 1em plus 0.5em minus 0.4em\relax IEEE, 2012, pp.
  629--632.

\bibitem{gruenewald2021tira}
E.~Grünewald, P.~Wille, F.~Pallas, M.~C. Borges, and M.-R. Ulbricht, ``{TIRA}:
  An {Open\-API} extension and toolbox for {GDPR} transparency in restful
  architectures,'' in \emph{2021 IEEE European Symposium on Security and
  Privacy Workshops (EuroS\&PW)}.\hskip 1em plus 0.5em minus 0.4em\relax IEEE
  Computer Society, 2021.

\bibitem{gruenewald2022devprivops}
E.~Grünewald, ``{Cloud Native Privacy Engineering through DevPrivOps},'' in
  \emph{Privacy and Identity Management. IFIP International Summer School,
  Esch-sur-Alzette, 2021}.\hskip 1em plus 0.5em minus 0.4em\relax Cham:
  Springer International Publishing, 2022. doi: 10.1007/978-3-030-99100-5\_10

\bibitem{cncf}
\BIBentryALTinterwordspacing
{Cloud Native Computing Foundation}. [Online]. Available:
  \url{https://github.com/cncf/toc/blob/main/DEFINITION.md}
\BIBentrySTDinterwordspacing

\bibitem{dragoni2017microservices}
N.~Dragoni, S.~Giallorenzo, A.~L. Lafuente, M.~Mazzara, F.~Montesi,
  R.~Mustafin, and L.~Safina, ``Microservices: yesterday, today, and
  tomorrow,'' \emph{Present and ulterior software engineering}, pp. 195--216,
  2017.

\bibitem{cloudnative}
D.~Gannon, R.~Barga, and N.~Sundaresan, ``Cloud-native applications,''
  \emph{IEEE Cloud Computing}, vol.~4, no.~5, pp. 16--21, 2017. doi:
  10.1109/MCC.2017.4250939

\bibitem{calcote2019istio}
L.~Calcote and Z.~Butcher, \emph{Istio: Up and running: Using a service mesh to
  connect, secure, control, and observe}.\hskip 1em plus 0.5em minus
  0.4em\relax O'Reilly Media, 2019.

\bibitem{li2019service}
W.~Li, Y.~Lemieux, J.~Gao, Z.~Zhao, and Y.~Han, ``Service mesh: Challenges,
  state of the art, and future research opportunities,'' in \emph{2019 IEEE
  International Conference on Service-Oriented System Engineering
  (SOSE)}.\hskip 1em plus 0.5em minus 0.4em\relax IEEE, 2019, pp. 122--1225.

\bibitem{chandramouli2020building}
R.~Chandramouli and Z.~Butcher, ``Building secure microservices-based
  applications using service-mesh architecture,'' \emph{NIST Special
  Publication}, vol. 800, p. 204A, 2020.

\bibitem{art29transparency}
\BIBentryALTinterwordspacing
{Article 29 Data Protection Working Party}, ``Guidelines on transparency under
  regulation 2016/679 -- wp260,'' 2018. [Online]. Available:
  \url{https://ec.europa.eu/newsroom/article29/redirection/document/51025}
\BIBentrySTDinterwordspacing

\bibitem{tilt}
E.~Grünewald and F.~Pallas, ``{TILT}: A {GDPR}-aligned transparency
  information language and toolkit for practical privacy engineering,'' in
  \emph{Proceedings of the 2021 Conference on Fairness, Accountability, and
  Transparency}, ser. FAccT '21.\hskip 1em plus 0.5em minus 0.4em\relax New
  York, NY, USA: Association for Computing Machinery, 2021. doi:
  10.1145/3442188.3445925

\bibitem{ReidenbergDisagreeablePrivacyPolicies2015}
J.~R. Reidenberg, T.~Breaux, L.~F. Cranor, B.~French, A.~Grannis, J.~T. Graves,
  F.~Liu, A.~McDonald, T.~B. Norton, and R.~Ramanath, ``Disagreeable {{Privacy
  Policies}}: {{Mismatches}} between {{Meaning}} and {{Users}}'
  {{Understanding}},'' \emph{Berkeley Technology Law Journal}, vol.~30, p.~39,
  2015.

\bibitem{Linden2018}
\BIBentryALTinterwordspacing
T.~Linden, R.~Khandelwal, H.~Harkous, and K.~Fawaz, ``The privacy policy
  landscape after the {GDPR},'' 2018. [Online]. Available:
  \url{http://arxiv.org/abs/1809.08396}
\BIBentrySTDinterwordspacing

\bibitem{rudolph2018}
M.~Rudolph, D.~Feth, and S.~Polst, ``Why users ignore privacy policies -- a
  survey and intention model for explaining user privacy behavior,'' in
  \emph{Human-Computer Interaction. Theories, Methods, and Human Issues}.\hskip
  1em plus 0.5em minus 0.4em\relax Cham: Springer, 2018. ISBN 978-3-319-91238-7

\bibitem{gruenewaldEnablingVersatile}
E.~Grünewald, J.~M. Halkenhäußer, N.~Leschke, J.~Washington, C.~Paupini, and
  F.~Pallas, ``Enabling versatile privacy interfaces using machine-readable
  transparency information,'' in \emph{Privacy Symposium 2023}, S.~Schiffner,
  A.~Q. Rodriguez, and S.~Ziegler, Eds.\hskip 1em plus 0.5em minus 0.4em\relax
  Springer International Publishing, 2023.

\bibitem{pribots}
\BIBentryALTinterwordspacing
H.~Harkous, K.~Fawaz, K.~G. Shin, and K.~Aberer, ``Pribots: Conversational
  privacy with chatbots,'' in \emph{Twelfth Symposium on Usable Privacy and
  Security ({SOUPS} 2016)}.\hskip 1em plus 0.5em minus 0.4em\relax Denver, CO:
  {USENIX} Association, 2016. [Online]. Available:
  \url{https://www.usenix.org/conference/soups2016/workshop-program/wfpn/presentation/harkous}
\BIBentrySTDinterwordspacing

\bibitem{holtz}
L.-E. Holtz, K.~Nocun, and M.~Hansen, ``Towards displaying privacy information
  with icons,'' in \emph{Privacy and Identity Management for Life},
  S.~Fischer-H{\"u}bner, P.~Duquenoy, M.~Hansen, R.~Leenes, and G.~Zhang,
  Eds.\hskip 1em plus 0.5em minus 0.4em\relax Berlin, Heidelberg: Springer
  Berlin Heidelberg, 2011. ISBN 978-3-642-20769-3 pp. 338--348.

\bibitem{habib2021icons}
H.~Habib, Y.~Zou, Y.~Yao, A.~Acquisti, L.~Cranor, J.~Reidenberg, N.~Sadeh, and
  F.~Schaub, ``Toggles, dollar signs, and triangles: How to (in)effectively
  convey privacy choices with icons and link texts,'' in \emph{Proceedings of
  the 2021 CHI Conference on Human Factors in Computing Systems}, ser. CHI
  '21.\hskip 1em plus 0.5em minus 0.4em\relax New York, NY, USA: Association
  for Computing Machinery, 2021. doi: 10.1145/3411764.3445387. ISBN
  9781450380966

\bibitem{gerl2019}
A.~Gerl and B.~Meier, ``The layered privacy language art. 12--14 {GDPR}
  extension--privacy enhancing user interfaces,'' \emph{Datenschutz und
  Datensicherheit-DuD}, vol.~43, no.~12, pp. 747--752, 2019.

\bibitem{bass2015devops}
L.~Bass, I.~Weber, and L.~Zhu, \emph{DevOps: A software architect's
  perspective}.\hskip 1em plus 0.5em minus 0.4em\relax Addison-Wesley
  Professional, 2015.

\bibitem{gurses_privacy_agile}
S.~Gürses and J.~van Hoboken, ``Privacy after the agile turn,'' in \emph{The
  Cambridge Handbook of Consumer Privacy}, ser. Cambridge Law Handbooks,
  E.~Selinger, J.~Polonetsky, and O.~Tene, Eds.\hskip 1em plus 0.5em minus
  0.4em\relax Cambridge University Press, 2018, p. 579–601.

\bibitem{kostova2020privacy}
B.~Kostova, S.~G{\"u}rses, and C.~Troncoso, ``Privacy engineering meets
  software engineering. on the challenges of engineering privacy bydesign,''
  \emph{arXiv:2007.08613}, 2020.

\bibitem{sion2020neverending}
L.~Sion, D.~V. Landuyt, and W.~Joosen, ``The never-ending story: On the need
  for continuous privacy impact assessment,'' in \emph{2020 IEEE European
  Symposium on Security and Privacy Workshops (EuroS PW)}, 2020. doi:
  10.1109/EuroSPW51379.2020.00049 pp. 314--317.

\bibitem{cnil}
\BIBentryALTinterwordspacing
{CNIL. Record of processing activities}. [Online]. Available:
  \url{https://www.cnil.fr/en/record-processing-activities}
\BIBentrySTDinterwordspacing

\bibitem{sion2021automated}
L.~Sion, D.~Van~Landuyt, K.~Yskout, S.~Verreydt, and W.~Joosen, ``Automated
  threat analysis and management in a continuous integration pipeline,'' in
  \emph{2021 IEEE Secure Development Conference (SecDev)}.\hskip 1em plus 0.5em
  minus 0.4em\relax IEEE, 2021, pp. 30--37.

\bibitem{data-track2016}
S.~Fischer-H{\"u}bner, J.~Angulo, F.~Karegar, and T.~Pulls, ``Transparency,
  privacy and trust -- technology for tracking and controlling my data
  disclosures: Does this work?'' in \emph{Trust Management X}, S.~M. Habib,
  J.~Vassileva, S.~Mauw, and M.~M{\"u}hlh{\"a}user, Eds.\hskip 1em plus 0.5em
  minus 0.4em\relax Cham: Springer International Publishing, 2016. ISBN
  978-3-319-41354-9 pp. 3--14.

\bibitem{kunz2020tracking}
I.~Kunz, V.~Casola, A.~Schneider, C.~Banse, and J.~Schütte, ``Towards tracking
  data flows in cloud architectures,'' in \emph{2020 IEEE 13th International
  Conference on Cloud Computing (CLOUD)}, 2020. doi:
  10.1109/CLOUD49709.2020.00066 pp. 445--452.

\bibitem{havur2020greater}
G.~Havur, M.~Vander~Sande, and S.~Kirrane, ``Greater control and transparency
  in personal data processing,'' 2020.

\bibitem{pasquier2016}
T.~Pasquier and D.~Eyers, ``Information flow audit for transparency and
  compliance in the handling of personal data,'' in \emph{2016 IEEE
  International Conference on Cloud Engineering Workshop (IC2EW)}, 2016. doi:
  10.1109/IC2EW.2016.29 pp. 112--117.

\bibitem{apiclarity}
\BIBentryALTinterwordspacing
``{APIClarity -- Reconstruct OpenAPI specifications from real-time workload
  traffic seamlessly}.'' [Online]. Available:
  \url{https://github.com/openclarity/apiclarity}
\BIBentrySTDinterwordspacing

\bibitem{redCastle}
F.~Pallas, J.~Legler, N.~Amslgruber, and E.~Grünewald, ``Redcastle:
  Practically applicable $k_s$-anonymity for iot streaming data at the edge in
  node-red,'' in \emph{Proceedings of the 8th International Workshop on
  Middleware and Applications for the Internet of Things}.\hskip 1em plus 0.5em
  minus 0.4em\relax New York: ACM, 2021. doi: 10.1145/3493369.3493601

\bibitem{sun2020security}
P.~Sun, ``Security and privacy protection in cloud computing: Discussions and
  challenges,'' \emph{Journal of Network and Computer Applications}, vol. 160,
  p. 102642, 2020.

\bibitem{sockshop}
\BIBentryALTinterwordspacing
{Sock Shop: A Microservice Demo Application}. [Online]. Available:
  \url{https://github.com/microservices-demo/microservices-demo}
\BIBentrySTDinterwordspacing

\bibitem{devopsRingu2016}
L.~Riungu-Kalliosaari, S.~M{\"a}kinen, L.~E. Lwakatare, J.~Tiihonen, and
  T.~M{\"a}nnist{\"o}, ``Devops adoption benefits and challenges in practice: A
  case study,'' in \emph{Product-Focused Software Process Improvement},
  P.~Abrahamsson, A.~Jedlitschka, A.~Nguyen~Duc, M.~Felderer, S.~Amasaki, and
  T.~Mikkonen, Eds.\hskip 1em plus 0.5em minus 0.4em\relax Cham: Springer
  International Publishing, 2016. ISBN 978-3-319-49094-6 pp. 590--597.

\bibitem{rapidErnst2019}
D.~Ernst, A.~Becker, and S.~Tai, ``Rapid canary assessment through proxying and
  two-stage load balancing,'' in \emph{2019 IEEE International Conference on
  Software Architecture Companion (ICSA-C)}, 2019. doi:
  10.1109/ICSA-C.2019.00028 pp. 116--122.

\bibitem{Finck_Biega_2021}
\BIBentryALTinterwordspacing
M.~Finck and A.~J. Biega, ``Reviving purpose limitation and data minimisation
  in data-driven systems,'' \emph{Technology and Regulation}, vol. 2021, p.
  44–61, Dec. 2021. doi: 10.26116/techreg.2021.004. [Online]. Available:
  \url{https://techreg.org/article/view/10986}
\BIBentrySTDinterwordspacing

\bibitem{finck2020they}
M.~Finck and F.~Pallas, ``They who must not be identified—distinguishing
  personal from non-personal data under the {GDPR},'' \emph{International Data
  Privacy Law}, vol.~10, no.~1, pp. 11--36, 2020.

\bibitem{gruenewald2022teiresias}
E.~Gr{\"u}newald and L.~Schurbert, ``Scalable discovery and continuous
  inventory of personal data at rest in cloud native systems,'' in
  \emph{Service-Oriented Computing}, J.~Troya, B.~Medjahed, M.~Piattini,
  L.~Yao, P.~Fern{\'a}ndez, and A.~Ruiz-Cort{\'e}s, Eds.\hskip 1em plus 0.5em
  minus 0.4em\relax Cham: Springer Nature Switzerland, 2022. ISBN
  978-3-031-20984-0 pp. 513--529.

\bibitem{sion-landuyt2021runtime}
L.~Sion, D.~V. Landuyt, and W.~Joosen, ``An overview of runtime data protection
  enforcement approaches,'' in \emph{2021 IEEE European Symposium on Security
  and Privacy Workshops (EuroS PW)}, 2021. doi: 10.1109/EuroSPW54576.2021.00044
  pp. 351--358.

\bibitem{pinisetty2018monitoring}
S.~Pinisetty, T.~Antignac, D.~Sands, and G.~Schneider, ``Monitoring data
  minimisation,'' \emph{arXiv:1801.02484}, 2018.

\bibitem{casola2015security}
V.~Casola, A.~De~Benedictis, and M.~Rak, ``Security monitoring in the cloud: an
  sla-based approach,'' in \emph{2015 10th International Conference on
  Availability, Reliability and Security}.\hskip 1em plus 0.5em minus
  0.4em\relax IEEE, 2015, pp. 749--755.

\bibitem{runtime2019}
J.~Happa, N.~Moffat, M.~Goldsmith, and S.~Creese, \emph{Run-Time Monitoring of
  Data-Handling Violations: Methods and Protocols}, 01 2019, pp. 213--232. ISBN
  978-1-4939-9041-2

\bibitem{angulo2015usable}
J.~Angulo, S.~Fischer-H{\"u}bner, T.~Pulls, and E.~W{\"a}stlund, ``Usable
  transparency with the data track: a tool for visualizing data disclosures,''
  in \emph{Proceedings of the 33rd Annual ACM Conference Extended Abstracts on
  Human Factors in Computing Systems}, 2015, pp. 1803--1808.

\bibitem{istio_bench}
\BIBentryALTinterwordspacing
{Istio Performance Benchmarking}. [Online]. Available:
  \url{https://github.com/istio/tools/tree/master/perf/benchmark}
\BIBentrySTDinterwordspacing

\bibitem{fortio}
\BIBentryALTinterwordspacing
{Fortio}. [Online]. Available: \url{https://github.com/fortio/fortio}
\BIBentrySTDinterwordspacing

\bibitem{nighthawk}
\BIBentryALTinterwordspacing
{Nighthawk}. [Online]. Available: \url{https://github.com/envoyproxy/nighthawk}
\BIBentrySTDinterwordspacing

\end{thebibliography}

\end{document}